\documentclass[prl,twocolumn,10pt,aps,nofootinbib,superscriptaddress,showkeys,showpacs]{revtex4-1}
\usepackage{amsmath,amsfonts,amssymb,amstext,amscd,amsthm,dsfont}
\usepackage{physics}
\usepackage{color}
\usepackage{soul,xcolor}
\usepackage{gensymb}
\usepackage[normalem]{ulem}

\usepackage[T1]{fontenc}

\usepackage{graphicx}

\usepackage{braket}
\usepackage{placeins}
\DeclareMathOperator{\spn}{span}

\begin{document}

\title{Non-Markovian Dynamics of Discrete-Time Quantum Walks}

\author{Subhashish Banerjee}
\email{subhashish@iitj.ac.in}
\affiliation{Indian Institute of Technology Jodhpur, Jodhpur 342011, India}

\author{N. Pradeep Kumar}
\email{pradeep5elangovan@gmail.com}
\affiliation{Indian Institute of Technology Jodhpur, Jodhpur 342011, India}

\author{R. Srikanth}
\email{srik@poornaprajna.org}
\affiliation{Poornaprajna Institute of Scientific Research,
Bangalore 560 080, India}

\author{Vinayak Jagadish} 
\email{vinayak.ukzn@gmail.com, JagadishV@ukzn.ac.za}

\author{Francesco Petruccione}
\email{petruccione@ukzn.ac.za}

\affiliation{Quantum Research Group, School  of Chemistry and Physics,
  University of KwaZulu-Natal, Durban 4001, South Africa, and National
  Institute  for Theoretical  Physics  (NITheP), KwaZulu-Natal,  South
  Africa}

\begin{abstract}
In  the case  of the  discrete time  coined quantum  walk the  reduced
dynamics of  the coin shows  non-Markovian recurrence features  due to
information back-flow  from the position  degree of freedom.   Here we
study how this  non-Markovian behavior is modified in  the presence of
open system dynamics.  In the process, we obtain  useful insights into
the nature of  non-Markovian physics.  In particular, we  show that in
the case  of (non-Markovian) random  telegraph noise (RTN),  a further
discernbile recurrence feature is  present in the dynamics.  Moreover,
this feature  is correlated with  the localization of the  walker.  On
the other  hand, no additional  recurruence feature appears  for other
non-Markovian types of noise (Ornstein-Uhlenbeck and Power Law noise).
We  propose  a  power  spectral  method  for  comparing  the  relative
strengths of the non-Markovian component due to the external noise and
that due to the internal position degree of freedom.
\end{abstract}
\pacs{03.65.Yz,03.67.-a}
\maketitle 


\textit{Introduction  ---}  Discrete-time  quantum Walk  (DTQW)  is  a
quantum analogue of ``classical random  walk'' (CRW) and describes the
evolution of a quantum particle  on a given topological structure. The
simplest instance  of DTQW is  that of a quantum system  with two
levels  translating  on  a  one dimensional  discrete  position  space
\cite{ambainis2001one,CBR10}, a toplogy which we use in this work. Any
practical implementation  of quantum walk demands  taking into account
the effects of the ambient  environment, resulting in the phenomena of
decoherence       and        dissipation       \cite{breuer2002theory,
  chandashekar2007symmetries, BRC+08}.


In the \textit{Markovian} regime, the environmental time scale is much
smaller   than   the   system   time  scale   \cite{VA17},   and   the
\textit{back-action}  by  system  on  the  environment,  in  terms  of
generating   system-environment  entanglement,   is  negligible.    In
contrast,  non-Markovian noise  features  backaction, and  furthermore
also ``back-flow" of  information from the environment  to the system,
which can  show up  as a  recurrence in  the correlations  between two
initial system states.  A particular  manifestation of recurrence is a
resonance  like  phenomena,  and   is  responsible  for  the  Anderson
localization observed in quantum walks \cite{PVB+17}.

With the  advancement of technologies,  one is  now able to  go beyond
Markovian phenomena and enter into  the non-Markovian regime, which we
undertake   in    this   work.    Unlike   previous    approaches   to
non-Markovianity,  our  approach  will distinguish  between  different
sources  of non-Markovianity,  in particular,  non-Markovian backflow.
As  a   concrete  application  of   our  approach,  we   study  coined
discrete-time   quantum  walk   (DTQW)   on  a   line,  subjected   to
\mbox{(non-)Markovian}  dynamics.   This   is  especially  interesting
because  the  reduced dynamics  of  the  coin manifests  non-Markovian
behavior due to the ``endemic'' source given by the position degree of
freedom \cite{Hinarejos2014}.  Our method will be able to disambiguate
the  non-Markovianity of  such an  endemic  origin versus  one due  to
environmental  decoherence. Here  we  make use  of  a local  dephasing
non-Markovian  noise model  \cite{daffer2004depolarizing} modelled  on
the random  telegraph noise (RTN)  process \cite{rice1944} as  well as
the            modified             Ornstein-Uhlenbeck            (OU)
\cite{uhlenbeck1930theory,yueberly2010} and the  power law noise (PLN)
\cite{KJ11}.

Localization, which  may be considered  as an aspect  of non-Markovian
backflow, was observed in \cite{benedetti2016non}, in the context of a
one dimensional continuous  time quantum walk, under  the influence of
RTN noise, in the presence of disorder. This behavior is also observed
here, in  the context of  DTQW on a line,  under the influence  of RTN
noise, which we report elsewhere \cite{PVB+17}.  Also observed are the
revival  of quantum  correlations in  the transition  from quantum  to
classical random walks, under the considered non-Markovian noise.

The  essential  ingredients of  DTQW  are  the \textit{coin}  and  the
\textit{position}   \cite{kempe2003quantum},   which   describes   the
internal   and  external   degrees   of  freedom   of  the   particle,
respectively. The  state of total  system is described by  the Hilbert
space $\mathcal{H}_w=\mathcal{H}_c \otimes  \mathcal{H}_p$ where $\spn
\mathcal{H}_c  =  \{\ket{0},  \ket{1}\}$  and  $\spn  \mathcal{H}_p  =
\{\ket{i}\}$, $i  \in \mathbb{Z}$  representing the number  of lattice
sites available to the walker. To  implement the DTQW, we initialize a
quantum  state $\rho$  and evolve  it using  the coin  and conditional
shift (i.e., translation in a  spatial dimension) operators.  The coin
operator  is usually  a two  dimensional rotation  matrix.  The  shift
operator  $\hat{S}$ that  translates the  particle to  either left  or
right is conditioned on the outcome of the coin operator.  The general
form  of the  shift operator  is given  as, $\hat{S}  = \ket{0}\bra{0}
\otimes \sum_{i\in\mathbb{Z}}  \ket{i-1}\bra{i}+\ket{1}\bra{1} \otimes
\sum_{i\in\mathbb{Z}}\ket{i+1}\bra{i}$.     $\rho(t)    =    \hat{W}^t
\rho(0)\hat{W}^{t\dagger}$, $\hat{W}^t$ is the  walk operator which is
a combination of $\hat{C}$ and $\hat{S}$ applied $t$ times.

\textit{Non-Markovian Noise ---} 
We will denote by $\Omega(t)$ the random variable describing the noise
fluctuation  in each  of the  three cases  and by  $M$ the  mean.  The
autocorrelation functions  and the  corresponding Kraus  operators are
summarised in table \ref{tab:noiseprop}.
\begin{widetext}
	\begin{table*}[htbp]
		\begin{tabular}{|c||c||c|}
			\hline
			Noise&Autocorrelation Function&Kraus Operators $(K_n)$\\
			\hline\hline
			RTN & $\displaystyle
			M[\Omega_i(t),\Omega_j(s)] = \delta_{ij} a^2 e^{-|t-s|/ \tau}$ &  $K_1    =    \sqrt{[1+\Lambda(\nu)]/2}I$,   $K_2    =
			\sqrt{[1-\Lambda(\nu)]/2}\sigma_3$   \\
			OUN & $\displaystyle
			M[\Omega(t),\Omega(s)]   =  \frac{\Gamma}{\gamma}  e^{-\gamma|t-s|}$ &  $K_1  =
			\ket{0}\bra{0}+p\ket{1}\bra{1}$, $K_2 = \sqrt{1-p^2}\ket{1}\bra{1}$   \\
			PLN & $\displaystyle  M[\Omega(t),\Omega(s)]  =
			\frac{1}{2}(\alpha-1)\alpha\Gamma\frac{1}{(\gamma|t-s|+1)^\alpha}$
			&  $K_1    =
			\ket{0}\bra{0}+q\ket{1}\bra{1}$, $K_2          =
			\sqrt{1-q^2}\ket{1}\bra{1}$  \\
			\hline
		\end{tabular}
		\caption{Autocorrelation functions and Kraus operators
                  for   three  types   of   noise.   $\Lambda(\nu)   =
                  e^{-\nu}[\text{cos}(\nu\mu)+\text{sin}(\nu\mu)/\mu],$
                  represents  the   damped  harmonic   function  which
                  encodes   both  the   Markovian  and   non-Markovian
                  behaviour of  the qubit, where $\displaystyle  \mu =
                  \sqrt{(2a/\gamma)^2-1}$  is  the  frequency  of  the
                  harmonic  oscillators  and  $\nu=\gamma  t$  is  the
                  dimensionless time.  For OU, $\gamma$  specifies the
                  noise  bandwidth  and   $\Gamma$  is  the  effective
                  relaxation     time.     $p     \equiv    p(t)     =
                  \text{exp}[-\frac{\Gamma}{2}
                    \{t+\frac{1}{\gamma}(e^{-\gamma         t}-1)\}]$.
                  $\alpha$ is some  real number and for  the $\alpha =
                  3$   case,   $\displaystyle    q   \equiv   q(t)   =
                  \text{exp}(-\frac{0.5t(t\gamma+2)\Gamma\gamma}
{(t\gamma+1)^2})$. }
		\label{tab:noiseprop}
	\end{table*}
\end{widetext}

In RTN,  $\gamma \equiv 1/2\tau$, $\tau$  being the time scale in which the
RTN noise changes its  phase.  The function $\Lambda(\nu)$ corresponds
to two  regimes; the purely  damping regime, with  Markovian behavior,
where  $2a/\gamma<1$,  and  damped  oscillations,  with  non-Markovian
behaviour, for  $2a/\gamma>1$, $a$  having the significance  of the
strength of  the system-environment coupling. The  regime of ``minimal
non-Markovian'' corresponds  to $2a/\gamma=1$, for which  $\mu=0$.  For the modified 
OU noise,  henceforth referred to as OUN, $\gamma^{-1}$   =  $\tau_c$,   where  $\tau_c$  is   the  finite
correlation time of the  environment (Table \ref{tab:noiseprop}).  In
Table  \ref{tab:noiseprop}, we also indicate properties  of  power-law  noise
(PLN), under which the DTQW behaves similar to the OUN \cite{PVB+17}.

A key feature of Markovian (memoryless) open-system dynamics $\Lambda$
is that  given two  distinct  states  $\rho$ and  $\sigma$, distance measures $\Delta$ (such as relative entropy or trace distance) satisfy $\mathfrak{D}[\Lambda(\rho),\Lambda(\sigma)] \le
\mathfrak{D}[\rho,\sigma]$, while  correlation measures $\mathfrak{C}$ (such as fidelity or mutual information)  satisfy $\mathfrak{C}[\Lambda(\rho),\Lambda(\sigma)]                       \ge \mathfrak{C}[\rho,\sigma]$.  By  contrast, non-Markovian  dynamics can violate the above  monotonicity property. In this work, we will use trace-distance  (TD) based  indicators,  and correlate  it with  other
features, such as oscillations in walk variance.  Non-Markovianity has
been  also  been  studied  using  fidelity  \cite{rajagopal2010kraus},
relative entropy \cite{DRS11} and mutual information \cite{PVB+17}.

\textit{Anderson Localization  ---} In  his seminal work  on transport
properties of  particles in a  random media, Anderson showed  that the
systems with  quenched disorder exibits the  phenomena of localization
\cite{anderson1958loc}.   In  the   quantum  walk   senario,  Anderson
Localization (AL) has  been studied extensively in  which the disorder
is   introduced    either   via   broken   links    in   the   lattice
\cite{benedetti2016non}   or  by   randomizing   the  coin   operation
\cite{chandru2012disorder}.  AL  can  be  interpreted  as  a  $memory$
property  of the  particle, as  it remembers  and localizies  near its
initial position when it is coupled to a disorded system. Here we show
that the AL phase can be observed in the non-Markovian (memory) regime
of RTN.

\begin{figure}
	\includegraphics[width=0.95\linewidth]{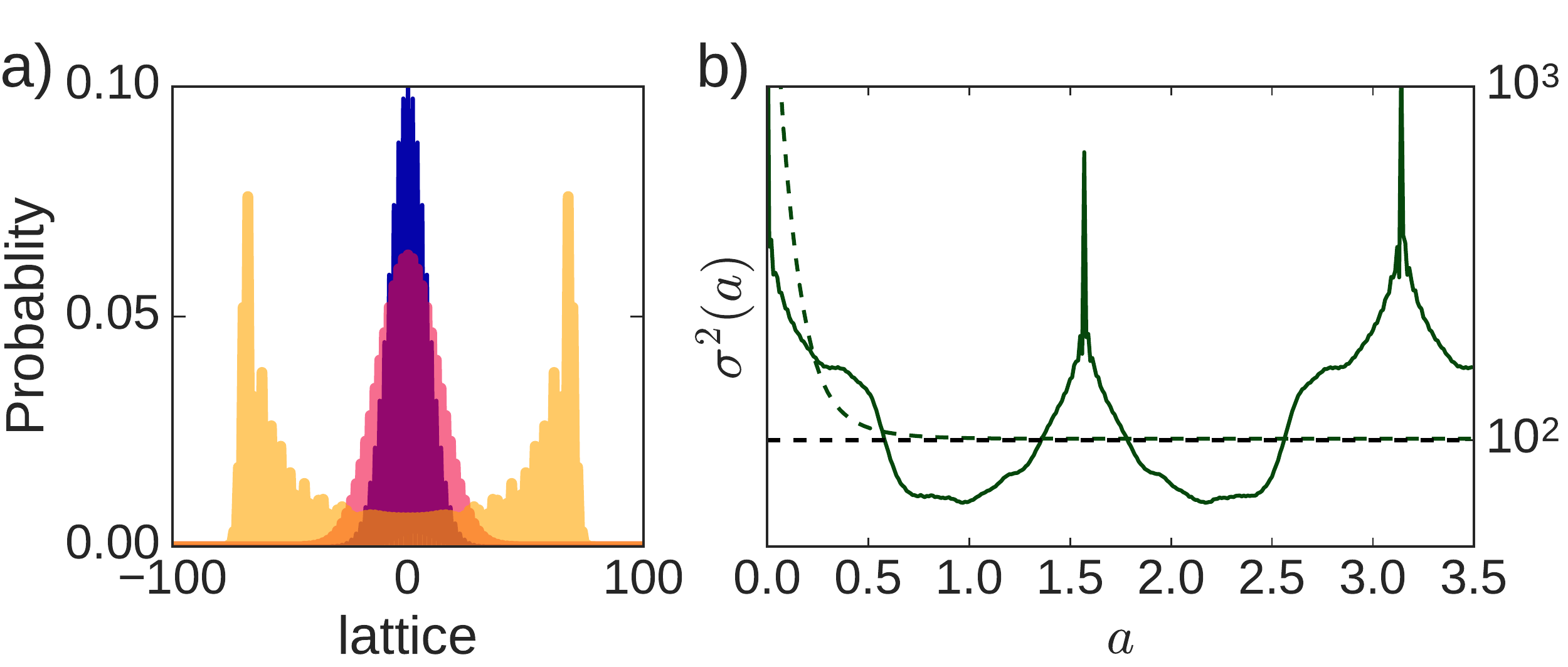}
	\caption{\small (Color online) (a) Probability distribution of
		the  QW in  the unitary  regime (bimodal),  Markovian (outer
		Gaussian,  $a=0.4,   \gamma=5$)  and   non-Markovian  (inner
		Gaussian, $a=1, \gamma=0.001$) regimes of RTN, using $t=100$
		steps.  (b) Log-linear plot of variance as a function of RTN
		amplitude $a$. The solid  (resp., dashed) line represent the
		corresponding  non-Markovian (resp.,  Markovian) cases  with
		$\gamma=0.001$  (resp.  $\gamma=5$).   The flat  dashed line
		represents  the classical  case.   We note  that the  former
		alternates between localization  (subclassical variance) and
		super-classical  variance.  The  Markovian case  shows plain
		decoherent  behavior, characterized  by  monotonic decay  of
		variance towards the classical value.}
		\label{fig:var_amp}
		\end{figure}

Figures   \ref{fig:var_amp}.(a)  and   (b)  displays   the  probablity
distribution and the corresponding variance as a function of the noise
amplitude $a$  of the quantum  walk under different noise  regimes. In
the absence  of noise the variance  of quantum walk shows  the typical
bimodal distribution, figure. \ref{fig:var_amp}.  (a) and the variance
evolves  quadratically  (ballistic  transport :  $var  \propto  t^2$).
Allowing for  the interaction  between RTN and  coin in  the Markovian
regime  ($\gamma=5.0$)  the  variance   decays  monotonically  to  the
classical random  walk limit  (diffusive transport: $var  \propto t$),
this is  shown as green  dashed line in  Figure \ref{fig:var_amp}.(b).
The non-Markovian regime  of RTN leads to interesting  features in the
quantum walk dynamics.   By tuning the correlation  time and amplitude
of RTN ($\gamma=0.001, a=1.0$) we can  observe AL phase, this is shown
as  solid line  in Figure  \ref{fig:var_amp}.(b). In  addition to  AL,
simply by  tuning the noise amplitude  $a$ of RTN we  observe that the
quantum  walk alternates  between three  different phases  namely, the
ballstic, diffusive and localization. In contrast to RTN, AL is absent
in both OUN and PLN.

\textit{Distinguishing  non-Markovian features  of  noise using  Trace distance ---} Trace
Distance    (TD)    \cite{laine2010measure}    is   a    measure    of
distinguishability between two states,  defined as $D(\rho_1,\rho_2) =
\frac{1}{2}Tr \norm{\rho_1-\rho_2}$, where  $\norm{A}$ is the operator
norm given by $\sqrt{A^\dag{A}}$.   For non-Markovian processes, owing
to the  backflow of  information from the  environment to  the system,
there could be an increase in the distinguishability, causing a deviation
from     the     monotonic     decrease     of     $D(\Phi(t)[\rho_1],
\Phi(t)[\rho_2(t)])$,  where  $\Phi(t)$  is the  noise  superoperator.
This idea has been exploited  in an effort to witness non-Markovianity
in \cite{breuer2009measure,breuer2016colloquium}. Another measure of non-Markovianity,
introduced in recent times, makes use of the deviation from CP of the 
intermediate  dynamics \cite{RHP10}.

\begin{figure}
	\includegraphics[width=\linewidth]{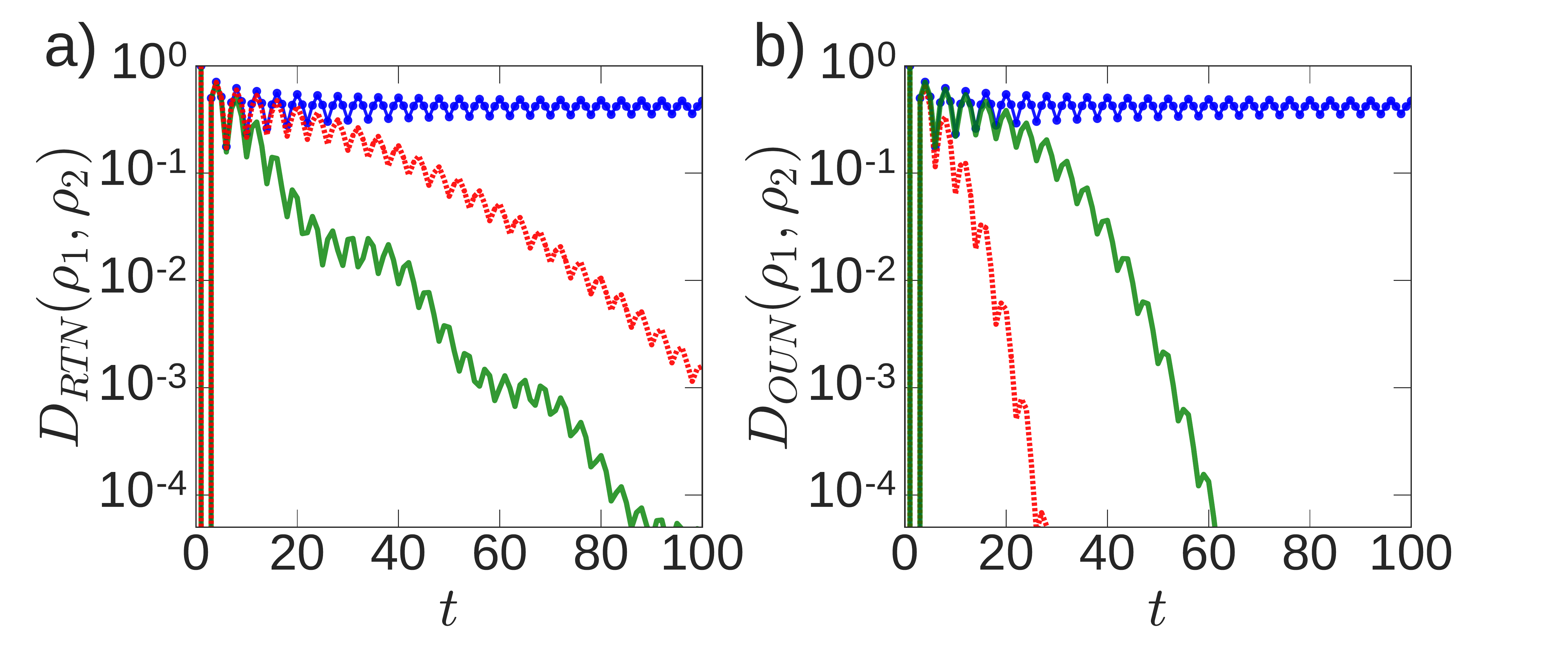}
	\caption{(Color  online)  Plot  of  TD  evolution,  under  the
          influence of RTN and OUN, with respect to the number of walk
          steps  $t$.   (a)  RTN:  The  trace  distance  plot  in  the
          noiseless quantum walk (top curve), in the Markovian (middle
          curve; $\gamma=1, a=0.05$)  and non-Markovian regime (bottom
          curve; $\gamma=0.001,  a=0.05$).  The pure walk  case, i.e.,
          the  QW in  the absence  of an  external noise,  shows rapid
          recurrences   due   to   interaction   with   the   position
          ``environment''  (primary  component  of  non-Markoviantiy),
          while the bottom curve  shows an additional oscillatory term
          (secondary component of  non-Markoviantiy) attributed to the
          non-Markovianity  in  the  environment-induced  decoherence.
          (b) OUN: In the Non-Markovian regime ($\gamma = 0.01, \Gamma
          = 0.1$), TD decays  without the additional recurrent feature
          seen  in the  RTN case  in  relation to  the Markovian  case
          ($\gamma =  1.0, \Gamma = 0.1$).   PLN is similar to  OUN in
          this respect.}
	\label{fig:RTN_td}
\end{figure}

Initially,  we  consider  the  noiseless (unitary)  evolution  of  the
quantum walk.   To study the  reduced dynamics  of the coin  state, we
compute  the  trace  distance  by  initializing  the  quantum  states,
$\ket{0}\pm\ket{1} /  \sqrt{2}$.  Since  the evolution is  governed by
unitary dynamics  the overall  evolution of  the quantum  walk remains
unitary and hence the TD is preserved. However, TD between the reduced
coin states undergoes  a high frequency oscillation, as  seen from the
top plots  in Figs \ref{fig:RTN_td}.(a)  and (b).  In  accordance with
the  criterion   for  non-Markovianity  in  terms   of  TD  recurrence
\cite{Hinarejos2014},  such   TD  oscillations  are  a   signature  of
non-Markovian behavior.  It is thus  important to note that we observe
non-Markovian behavior in the coin {\it even} in the unitary dynamics,
owing to the evolution of its entanglement with the position degree of
freedom.   Not  surprisingly,  the   oscillations  are  large  at  the
beginning, when  the position  dimensionality is  small, and  then die
out, as the position dimensionality gets larger.

\textit{Power spectral analysis --- }
In addition to the oscillations  present in the noiseless evolution of
quantum walk, a  further backflow or recurrence  structure arises when
the  coin  is  exposed  to  an  external noise  such  as  RTN  in  the
non-Markovian  regime.  In  order to  disambiguate these  two distinct
sources of non-Markovianity, we compute the power spectrum of the time
evolution $D(\rho_1(t),\rho_2(t))$ of correlation-like quantities such
as  trace   distance  or   mutual  information,  minus   the  function
$\delta_B(t)$, the  Monotonically Falling Best Fit (MFBF) function, which is
the   monotonically    falling   function    that   is    closest   to
$D(\rho_1(t),\rho_2(t))$,  according to  a suitable  distance measure. This  allows us  to usually  locate in  the frequency  domain the different sources of the backflow  aspect of non-Markovianity. In this case, the position  degree of freedom serves as one  source, while RTN producing environment serves as the other source.

The  power spectrum  of the  evolution  is computed  for time  $N=100$
steps,  as the  absolute  squared of  the  Discrete Fourier  Transform
(DFT),   i.e.,   $\mathcal{S}(k)=|\sum_{n=0}^{N-1}  x_n   e^{-2\pi   i
	kn/N}|^2$,  where $x_n$  is TD  or  mutual information  in the  time
domain at step $n$.

\textit{Case   of  RTN ---}   Figure  \ref{fig:RTN_td}.(a)   depicts  time
evolution of  the DTQW in the  noiseless case (top plot),  with RTN in
the Markovian regime  (middle plot) and with RTN  in the non-Markovian
regime  (bottom  plot).   The high-frequency  ``primary''  oscillation
(which corresponds to non-Markovianity according to the Breuer measure
\cite{breuer2009measure}) in the top plot is due to the interaction of
the position  and coin  degrees of freedom  \cite{Hinarejos2014}. Note
the  ring-down of  the oscillations.   In the  middle plot,  Markovian
decoherence  with RTN  is  seen to  cause a  monotonic  fall, with  an
overlay  of   the  position-induced   oscillation.  Finally,   in  the
bottom-most plot,  RTN in the  non-Markovian regime introduces  a new,
``secondary'', lower-frequency oscillation component.

\begin{figure}
	\centering
	\includegraphics[width=\linewidth]{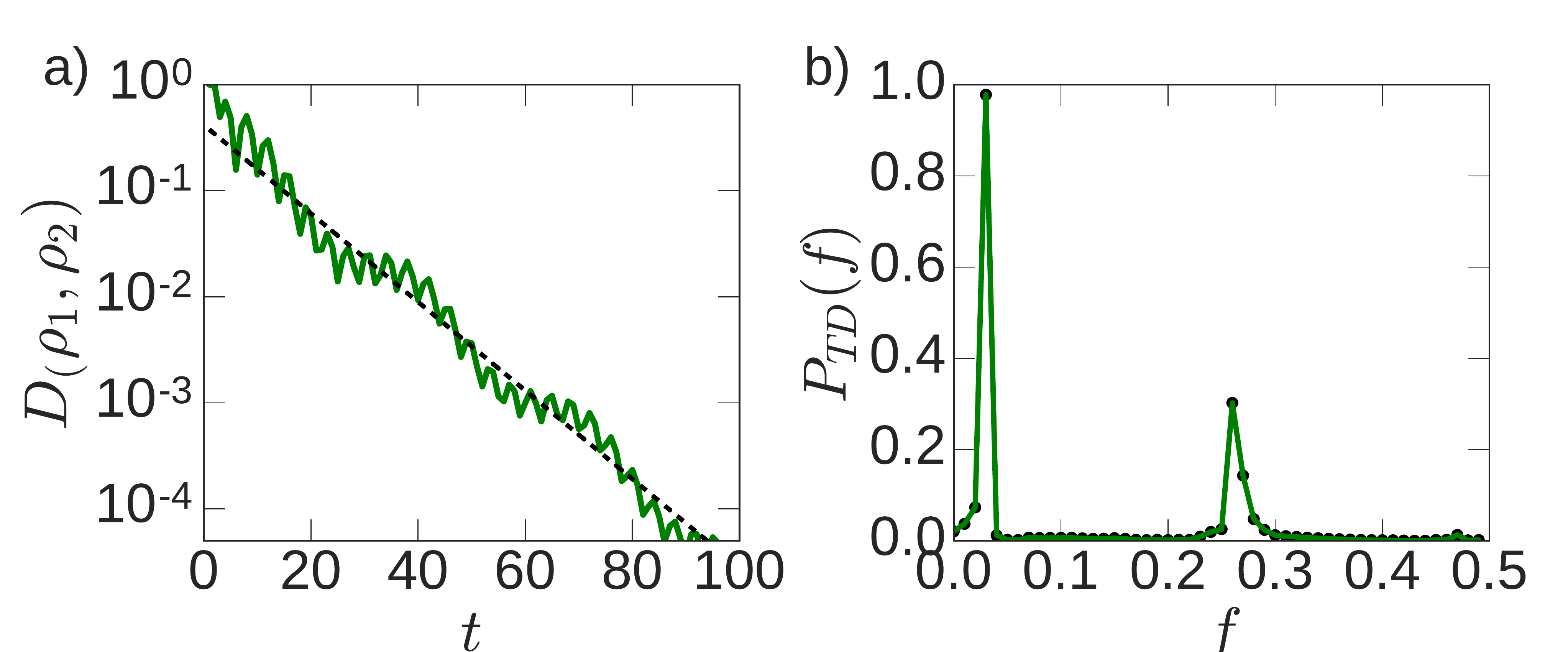}
	\caption{(a) Power  spectrum analysis  of the  two oscillatory
          components,  primary (due  to position  ``environment'') and
          secondary (due to RTN).  The bottom plot \ref{fig:RTN_td}(a)
          is subjected  to filtering  by subtracting  the straightline
          best fit to  the plot (which corresponds  to $\sim 0.4\times
          0.9^{-t}$).  (b)  The power  spectrum of the  filtered plot,
          with  respect to  the frequency  $f$, where  the frequencies
          correspond precisely  to the oscillations seen  in (a), with
          the one around $f=0.27$  (resp., $f=0.03$) corresponding the
          primary  (resp.,   secondary)  recurrence   component.   The
          importance of this  filtering approach is that  it allows us
          to  compare the  relative strengths  of the  two sources  of
          non-Markovianity (the  area under  the primary is  1.5 times
          more). In OUN or PLN,  the secondary recurrence component is
          not seen; just the diminution of the primary component.}
	\label{fig:Filtered_spectrum}
\end{figure}

Figure \ref{fig:RTN_td}.(b)  depicts the corresponding  situation with
OUN, where,the  secondary oscillatory component, due  to backflow from
the  external environment,  and as  depicted  in the  middle plot,  is
missing even  in the  non-Markovian regime.   Instead non-Markovianity
manifests as a backaction of the  system on the reservoir, producing a
slowing down of the decoherence  rate.  This phenomenon is essentially
due  to  the  low  bandwidth of  the  reservoir  frequency,  $\gamma$,
resulting in  a large  reservoir correlation time  in relation  to the
system correlation time.

In Figure  \ref{fig:Filtered_spectrum}, we present the  power spectrum
of the bottom  plot of Figure \ref{fig:RTN_td}(a), from  which we have
subtracted  the  MFBF.  The  rationale  is,  as  noted  earlier,  that
Markovian  dynamics  can only  generate  a  monotonic reduction  of  a
distance  measure.    Any  departure   from  monotonicity   should  be
attributed, then,  to non-Markovianity,  in particular  backflow.  The
filtered plot can  thus be considered as a measure  of the lower bound
on non-Markovian behavior.   In the present case,  the subtracted part
turns out to be a simple power  law (represented as a straight line in
the   log-normal  plot).    More  generally,   this  subtracted   part
corresponds  to a  problem of  monotonic curve  fitting, which  can be
formulated as  a semidefinite program \cite{PVB+17}.   The peak around
$f=0.27$  (resp.   $f=0.025$)  corresponds   to  the  primary  (resp.,
secondary) non-Markovian source, namely the position degree of freedom
(resp., the environment). The information backflow is a resonance like
phenomena, producing the secondary  peak.  Thus, our approach provides
a tool to disambiguate two sources of non-Markovian backflow.

\begin{figure}
	\centering
	\includegraphics[width=\linewidth]{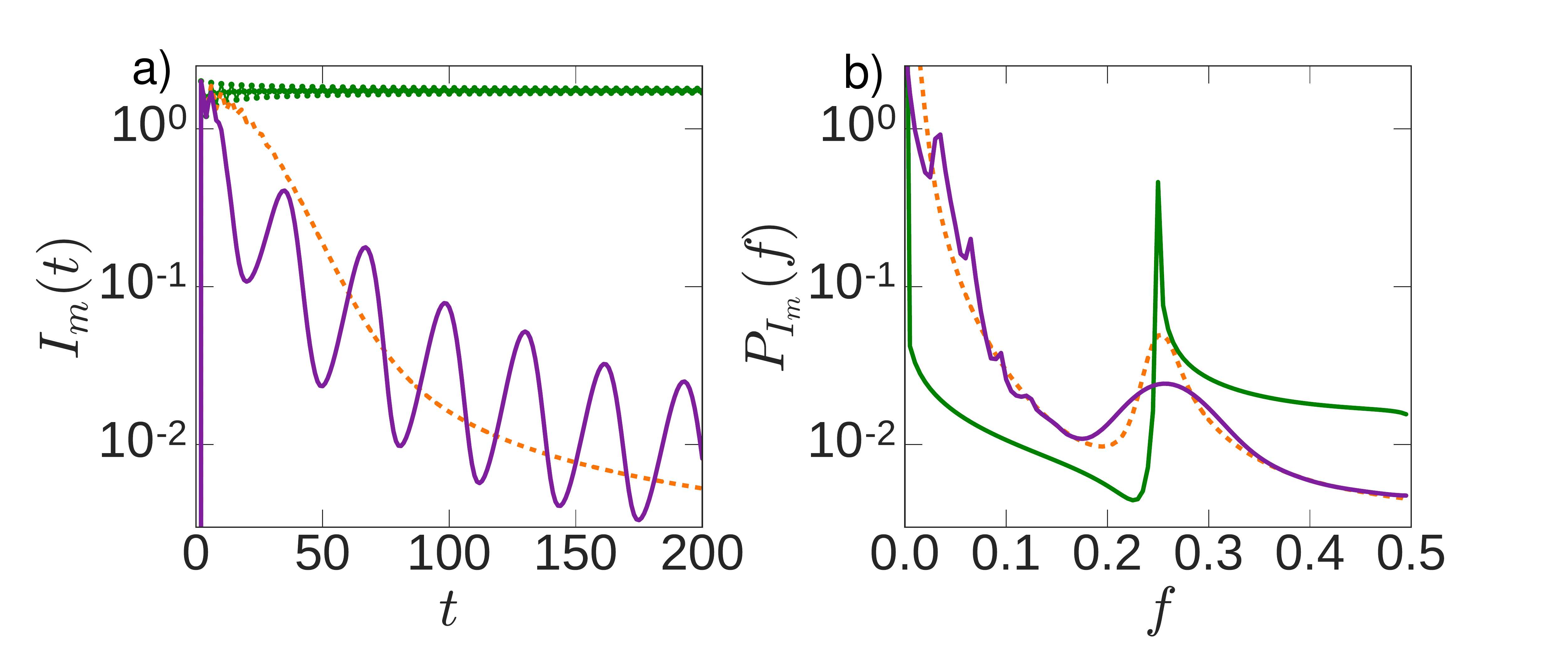}
	\caption{(Color online) Non-markovianity  studied using mutual
		information  (MI)  for  RTN  with  noise  amplitude  set  to
		$a=0.05$. (a)  The top curve (green) shows recurrence
		in the pure  QW case due to entanglement  with position; the
		monotonically  falling curve(orange) is  QW in  the
		Markovian regime ($a=0.05, \gamma=1$),  while the
		oscillating    curve   (purple)   represents    the
		non-Markovian  regime ($a=0.05,  \gamma=0.001$), manifesting
		both the primary oscillatory  component (seen in the unitary
		case  itself) and  the  second oscillatory  overlap (due  to
		RTN);  (b) The  power  spectrum corresponding  to the  three
		curves,  indicated  using the  same  plot  styles.  Note  the
		pronounced primary component  in the pure QW  case, with the
		secondary component showing  up as a series of  peaks in the
		power spectrum. But contrast this secondary part is not seen
		in the Markovian power spectrum.}
	\label{fig:RTN_MI}
\end{figure}

\textit{Detecting non-Markovianity using Mutual Information ---}
Quantum correlations as quantified by mutual information (MI) has been
used  to  quantify  non-Markovianity  \cite{luo2012quantifying}.   Let
$\rho_1$ and $\rho_2$ be the  density matrices representing the system
and  the  ancillary  state, respectively.   Given  density  operators
$\rho_1$    and     $\rho_2$,    their    mutual     information    is
$\mathcal{I}_m(\rho)  = S(\rho_1)+S(\rho_2)-S(\rho_{12})$,  where S(.)
is  the   von  Neumann  entropy   $S(\rho)$:=-tr$\rho$log$_2$  $\rho$.
Similar  to the  TD measure,  MI  is also  a monotonically  decreasing
function when  the dynamics is Markovian.   Figure \ref{fig:RTN_MI}(a)
presents  the  MI  equivalent  of  Figure  \ref{fig:RTN_td}(a),  while
\ref{fig:RTN_MI}(b)  presents  the  power  spectrum of  the  plots  of
\ref{fig:RTN_MI}(a). We note the  primary peak around $f=0.27$.  Apart
from this, the spectrum of the Markovian noise is smooth, whereas that
of  the  non-Markovian  case  shows  signatures  of  secondary  peaks,
indicating non-Markovianity of environmental origin.  Note that the spectral filtering method can also be employed for MI by finding a suitable MFBF function.  

\textit{Conclusion ---}  Information backflow from the  environment to
the  system  is an  important  aspect  of non-Markovianity  absent  in
Markovian  dynamics.  In  a  system  such as  a  coined quantum  walk,
information backflow appears  in the reduced coin  system dynamics due
to    both   environmental    decoherence   and    the   coin-position
interaction. Here  we have developed  tools to disambiguate  these two
sources of  non-Markovianity.  We  identify backflow  by peaks  in the
power spectrum of time evolution of distance or nearness measures of a
pair of quantum states.  All  known measures of non-Markovian behavior
are incapable of making this distinction.  This work thus presents
novel  insights  into  the   nature  and  detection  of  non-Markovian
evolution.

\textit{Acknowledgments ---}  The work  of V.~J.  and F.~P.   is based
upon research supported by the South African Research Chair Initiative
of  the Department  of Science  and Technology  and National  Research
Foundation.

\bibliographystyle{apsrev4-1}
\bibliography{QW_NM}

\begin{thebibliography}{24}%
\makeatletter
\providecommand \@ifxundefined [1]{%
 \@ifx{#1\undefined}
}%
\providecommand \@ifnum [1]{%
 \ifnum #1\expandafter \@firstoftwo
 \else \expandafter \@secondoftwo
 \fi
}%
\providecommand \@ifx [1]{%
 \ifx #1\expandafter \@firstoftwo
 \else \expandafter \@secondoftwo
 \fi
}%
\providecommand \natexlab [1]{#1}%
\providecommand \enquote  [1]{``#1''}%
\providecommand \bibnamefont  [1]{#1}%
\providecommand \bibfnamefont [1]{#1}%
\providecommand \citenamefont [1]{#1}%
\providecommand \href@noop [0]{\@secondoftwo}%
\providecommand \href [0]{\begingroup \@sanitize@url \@href}%
\providecommand \@href[1]{\@@startlink{#1}\@@href}%
\providecommand \@@href[1]{\endgroup#1\@@endlink}%
\providecommand \@sanitize@url [0]{\catcode `\\12\catcode `\$12\catcode
  `\&12\catcode `\#12\catcode `\^12\catcode `\_12\catcode `\%12\relax}%
\providecommand \@@startlink[1]{}%
\providecommand \@@endlink[0]{}%
\providecommand \url  [0]{\begingroup\@sanitize@url \@url }%
\providecommand \@url [1]{\endgroup\@href {#1}{\urlprefix }}%
\providecommand \urlprefix  [0]{URL }%
\providecommand \Eprint [0]{\href }%
\providecommand \doibase [0]{http://dx.doi.org/}%
\providecommand \selectlanguage [0]{\@gobble}%
\providecommand \bibinfo  [0]{\@secondoftwo}%
\providecommand \bibfield  [0]{\@secondoftwo}%
\providecommand \translation [1]{[#1]}%
\providecommand \BibitemOpen [0]{}%
\providecommand \bibitemStop [0]{}%
\providecommand \bibitemNoStop [0]{.\EOS\space}%
\providecommand \EOS [0]{\spacefactor3000\relax}%
\providecommand \BibitemShut  [1]{\csname bibitem#1\endcsname}%
\let\auto@bib@innerbib\@empty
\bibitem [{\citenamefont {Ambainis}\ \emph {et~al.}(2001)\citenamefont
  {Ambainis}, \citenamefont {Bach}, \citenamefont {Nayak}, \citenamefont
  {Vishwanath},\ and\ \citenamefont {Watrous}}]{ambainis2001one}%
  \BibitemOpen
  \bibfield  {author} {\bibinfo {author} {\bibfnamefont {A.}~\bibnamefont
  {Ambainis}}, \bibinfo {author} {\bibfnamefont {E.}~\bibnamefont {Bach}},
  \bibinfo {author} {\bibfnamefont {A.}~\bibnamefont {Nayak}}, \bibinfo
  {author} {\bibfnamefont {A.}~\bibnamefont {Vishwanath}}, \ and\ \bibinfo
  {author} {\bibfnamefont {J.}~\bibnamefont {Watrous}},\ }in\ \href@noop {}
  {\emph {\bibinfo {booktitle} {Proceedings of the thirty-third annual ACM
  symposium on Theory of computing}}}\ (\bibinfo {organization} {ACM},\
  \bibinfo {year} {2001})\ pp.\ \bibinfo {pages} {37--49}\BibitemShut {NoStop}%
\bibitem [{\citenamefont {Chandrashekar}\ \emph {et~al.}(2010)\citenamefont
  {Chandrashekar}, \citenamefont {Banerjee},\ and\ \citenamefont
  {Srikanth}}]{CBR10}%
  \BibitemOpen
  \bibfield  {author} {\bibinfo {author} {\bibfnamefont {C.}~\bibnamefont
  {Chandrashekar}}, \bibinfo {author} {\bibfnamefont {S.}~\bibnamefont
  {Banerjee}}, \ and\ \bibinfo {author} {\bibfnamefont {R.}~\bibnamefont
  {Srikanth}},\ }\href@noop {} {\bibfield  {journal} {\bibinfo  {journal} {Phy.
  Rev. A}\ }\textbf {\bibinfo {volume} {81}},\ \bibinfo {pages} {062340}
  (\bibinfo {year} {2010})}\BibitemShut {NoStop}%
\bibitem [{\citenamefont {Breuer}\ and\ \citenamefont
  {Petruccione}(2002)}]{breuer2002theory}%
  \BibitemOpen
  \bibfield  {author} {\bibinfo {author} {\bibfnamefont {H.-P.}\ \bibnamefont
  {Breuer}}\ and\ \bibinfo {author} {\bibfnamefont {F.}~\bibnamefont
  {Petruccione}},\ }\href@noop {} {\emph {\bibinfo {title} {The theory of open
  quantum systems}}}\ (\bibinfo  {publisher} {Oxford University Press on
  Demand},\ \bibinfo {year} {2002})\BibitemShut {NoStop}%
\bibitem [{\citenamefont {Chandrashekar}\ \emph {et~al.}(2007)\citenamefont
  {Chandrashekar}, \citenamefont {Srikanth},\ and\ \citenamefont
  {Banerjee}}]{chandashekar2007symmetries}%
  \BibitemOpen
  \bibfield  {author} {\bibinfo {author} {\bibfnamefont {C.}~\bibnamefont
  {Chandrashekar}}, \bibinfo {author} {\bibfnamefont {R.}~\bibnamefont
  {Srikanth}}, \ and\ \bibinfo {author} {\bibfnamefont {S.}~\bibnamefont
  {Banerjee}},\ }\href@noop {} {\bibfield  {journal} {\bibinfo  {journal} {Phy.
  Rev. A}\ }\textbf {\bibinfo {volume} {76}},\ \bibinfo {pages} {022316}
  (\bibinfo {year} {2007})}\BibitemShut {NoStop}%
\bibitem [{\citenamefont {Banerjee}\ \emph {et~al.}(2008)\citenamefont
  {Banerjee}, \citenamefont {Srikanth}, \citenamefont {Chandrashekar},\ and\
  \citenamefont {Rungta}}]{BRC+08}%
  \BibitemOpen
  \bibfield  {author} {\bibinfo {author} {\bibfnamefont {S.}~\bibnamefont
  {Banerjee}}, \bibinfo {author} {\bibfnamefont {R.}~\bibnamefont {Srikanth}},
  \bibinfo {author} {\bibfnamefont {C.}~\bibnamefont {Chandrashekar}}, \ and\
  \bibinfo {author} {\bibfnamefont {P.}~\bibnamefont {Rungta}},\ }\href@noop {}
  {\bibfield  {journal} {\bibinfo  {journal} {Phys. Rev. A}\ }\textbf {\bibinfo
  {volume} {78}},\ \bibinfo {pages} {052316} (\bibinfo {year}
  {2008})}\BibitemShut {NoStop}%
\bibitem [{\citenamefont {de~Vega}\ and\ \citenamefont {Alonso}(2017)}]{VA17}%
  \BibitemOpen
  \bibfield  {author} {\bibinfo {author} {\bibfnamefont {I.}~\bibnamefont
  {de~Vega}}\ and\ \bibinfo {author} {\bibfnamefont {D.}~\bibnamefont
  {Alonso}},\ }\href {\doibase 10.1103/RevModPhys.89.015001} {\bibfield
  {journal} {\bibinfo  {journal} {Rev. Mod. Phys.}\ }\textbf {\bibinfo {volume}
  {89}},\ \bibinfo {pages} {015001} (\bibinfo {year} {2017})}\BibitemShut
  {NoStop}%
\bibitem [{\citenamefont {Pradeep~Kumar}\ \emph {et~al.}()\citenamefont
  {Pradeep~Kumar}, \citenamefont {Jagadish}, \citenamefont {Banerjee},
  \citenamefont {Srikanth},\ and\ \citenamefont {Petruccione}}]{PVB+17}%
  \BibitemOpen
  \bibfield  {author} {\bibinfo {author} {\bibfnamefont {N.}~\bibnamefont
  {Pradeep~Kumar}}, \bibinfo {author} {\bibfnamefont {V.}~\bibnamefont
  {Jagadish}}, \bibinfo {author} {\bibfnamefont {S.}~\bibnamefont {Banerjee}},
  \bibinfo {author} {\bibfnamefont {R.}~\bibnamefont {Srikanth}}, \ and\
  \bibinfo {author} {\bibfnamefont {F.}~\bibnamefont {Petruccione}},\
  }\href@noop {} {\bibinfo  {journal} {Draft under preparation}\ }\BibitemShut
  {NoStop}%
\bibitem [{\citenamefont {Hinarejos}\ \emph {et~al.}(2010)\citenamefont
  {Hinarejos}, \citenamefont {Franco}, \citenamefont {Romanelli},\ and\
  \citenamefont {Perez}}]{Hinarejos2014}%
  \BibitemOpen
\bibfield  {journal} {  }\bibfield  {author} {\bibinfo {author} {\bibfnamefont
  {M.}~\bibnamefont {Hinarejos}}, \bibinfo {author} {\bibfnamefont {C.~D.}\
  \bibnamefont {Franco}}, \bibinfo {author} {\bibfnamefont {A.}~\bibnamefont
  {Romanelli}}, \ and\ \bibinfo {author} {\bibfnamefont {A.}~\bibnamefont
  {Perez}},\ }\href@noop {} {\bibfield  {journal} {\bibinfo  {journal} {Phys.
  Rev. A}\ }\textbf {\bibinfo {volume} {81}},\ \bibinfo {pages} {014101}
  (\bibinfo {year} {2010})}\BibitemShut {NoStop}%
\bibitem [{\citenamefont {Daffer}\ \emph {et~al.}(2004)\citenamefont {Daffer},
  \citenamefont {W{\'o}dkiewicz}, \citenamefont {Cresser},\ and\ \citenamefont
  {McIver}}]{daffer2004depolarizing}%
  \BibitemOpen
  \bibfield  {author} {\bibinfo {author} {\bibfnamefont {S.}~\bibnamefont
  {Daffer}}, \bibinfo {author} {\bibfnamefont {K.}~\bibnamefont
  {W{\'o}dkiewicz}}, \bibinfo {author} {\bibfnamefont {J.~D.}\ \bibnamefont
  {Cresser}}, \ and\ \bibinfo {author} {\bibfnamefont {J.~K.}\ \bibnamefont
  {McIver}},\ }\href@noop {} {\bibfield  {journal} {\bibinfo  {journal} {Phys.
  Rev. A}\ }\textbf {\bibinfo {volume} {70}},\ \bibinfo {pages} {010304}
  (\bibinfo {year} {2004})}\BibitemShut {NoStop}%
\bibitem [{\citenamefont {Rice}()}]{rice1944}%
  \BibitemOpen
  \bibfield  {author} {\bibinfo {author} {\bibfnamefont {S.~O.}\ \bibnamefont
  {Rice}},\ }\href@noop {} {\bibinfo  {journal} {Stochastic Processes in
  Physics and Chemistry (Elsevier, Amsterdam, 1992)}\ }\BibitemShut {NoStop}%
\bibitem [{\citenamefont {Uhlenbeck}\ and\ \citenamefont
  {Ornstein}(1930)}]{uhlenbeck1930theory}%
  \BibitemOpen
\bibfield  {journal} {  }\bibfield  {author} {\bibinfo {author} {\bibfnamefont
  {G.~E.}\ \bibnamefont {Uhlenbeck}}\ and\ \bibinfo {author} {\bibfnamefont
  {L.~S.}\ \bibnamefont {Ornstein}},\ }\href@noop {} {\bibfield  {journal}
  {\bibinfo  {journal} {Phys. Rev}\ }\textbf {\bibinfo {volume} {36}},\
  \bibinfo {pages} {823} (\bibinfo {year} {1930})}\BibitemShut {NoStop}%
\bibitem [{\citenamefont {Yu}\ and\ \citenamefont
  {Eberly}(2010)}]{yueberly2010}%
  \BibitemOpen
  \bibfield  {author} {\bibinfo {author} {\bibfnamefont {T.}~\bibnamefont
  {Yu}}\ and\ \bibinfo {author} {\bibfnamefont {J.}~\bibnamefont {Eberly}},\
  }\href@noop {} {\bibfield  {journal} {\bibinfo  {journal} {Opt. Commun}\
  }\textbf {\bibinfo {volume} {283}},\ \bibinfo {pages} {676} (\bibinfo {year}
  {2010})}\BibitemShut {NoStop}%
\bibitem [{\citenamefont {Kendal}\ and\ \citenamefont
  {J\o{}rgensen}(2011)}]{KJ11}%
  \BibitemOpen
  \bibfield  {author} {\bibinfo {author} {\bibfnamefont {W.~S.}\ \bibnamefont
  {Kendal}}\ and\ \bibinfo {author} {\bibfnamefont {B.}~\bibnamefont
  {J\o{}rgensen}},\ }\href {\doibase 10.1103/PhysRevE.84.066120} {\bibfield
  {journal} {\bibinfo  {journal} {Phys. Rev. E}\ }\textbf {\bibinfo {volume}
  {84}},\ \bibinfo {pages} {066120} (\bibinfo {year} {2011})}\BibitemShut
  {NoStop}%
\bibitem [{\citenamefont {Benedetti}\ \emph {et~al.}(2016)\citenamefont
  {Benedetti}, \citenamefont {Buscemi}, \citenamefont {Bordone},\ and\
  \citenamefont {Paris}}]{benedetti2016non}%
  \BibitemOpen
  \bibfield  {author} {\bibinfo {author} {\bibfnamefont {C.}~\bibnamefont
  {Benedetti}}, \bibinfo {author} {\bibfnamefont {F.}~\bibnamefont {Buscemi}},
  \bibinfo {author} {\bibfnamefont {P.}~\bibnamefont {Bordone}}, \ and\
  \bibinfo {author} {\bibfnamefont {M.~G.}\ \bibnamefont {Paris}},\ }\href@noop
  {} {\bibfield  {journal} {\bibinfo  {journal} {Phys. Rev. A}\ }\textbf
  {\bibinfo {volume} {93}},\ \bibinfo {pages} {042313} (\bibinfo {year}
  {2016})}\BibitemShut {NoStop}%
\bibitem [{\citenamefont {Kempe}(2003)}]{kempe2003quantum}%
  \BibitemOpen
  \bibfield  {author} {\bibinfo {author} {\bibfnamefont {J.}~\bibnamefont
  {Kempe}},\ }\href@noop {} {\bibfield  {journal} {\bibinfo  {journal} {Cont.
  Phy}\ }\textbf {\bibinfo {volume} {44}},\ \bibinfo {pages} {307} (\bibinfo
  {year} {2003})}\BibitemShut {NoStop}%
\bibitem [{\citenamefont {Rajagopal}\ \emph {et~al.}(2010)\citenamefont
  {Rajagopal}, \citenamefont {Devi},\ and\ \citenamefont
  {Rendell}}]{rajagopal2010kraus}%
  \BibitemOpen
  \bibfield  {author} {\bibinfo {author} {\bibfnamefont {A.}~\bibnamefont
  {Rajagopal}}, \bibinfo {author} {\bibfnamefont {A.~U.}\ \bibnamefont {Devi}},
  \ and\ \bibinfo {author} {\bibfnamefont {R.}~\bibnamefont {Rendell}},\
  }\href@noop {} {\bibfield  {journal} {\bibinfo  {journal} {Phys. Rev. A}\
  }\textbf {\bibinfo {volume} {82}},\ \bibinfo {pages} {042107} (\bibinfo
  {year} {2010})}\BibitemShut {NoStop}%
\bibitem [{\citenamefont {Usha~Devi}\ \emph {et~al.}(2011)\citenamefont
  {Usha~Devi}, \citenamefont {Rajagopal},\ and\ \citenamefont {Sudha}}]{DRS11}%
  \BibitemOpen
  \bibfield  {author} {\bibinfo {author} {\bibfnamefont {A.~R.}\ \bibnamefont
  {Usha~Devi}}, \bibinfo {author} {\bibfnamefont {A.~K.}\ \bibnamefont
  {Rajagopal}}, \ and\ \bibinfo {author} {\bibnamefont {Sudha}},\ }\href
  {\doibase 10.1103/PhysRevA.83.022109} {\bibfield  {journal} {\bibinfo
  {journal} {Phys. Rev. A}\ }\textbf {\bibinfo {volume} {83}},\ \bibinfo
  {pages} {022109} (\bibinfo {year} {2011})}\BibitemShut {NoStop}%
\bibitem [{\citenamefont {Anderson}(1958)}]{anderson1958loc}%
  \BibitemOpen
  \bibfield  {author} {\bibinfo {author} {\bibfnamefont {P.~W.}\ \bibnamefont
  {Anderson}},\ }\href@noop {} {\bibfield  {journal} {\bibinfo  {journal}
  {Phys. Rev.}\ }\textbf {\bibinfo {volume} {109}},\ \bibinfo {pages} {1492}
  (\bibinfo {year} {1958})}\BibitemShut {NoStop}%
\bibitem [{\citenamefont {Chandrashekar}(2012)}]{chandru2012disorder}%
  \BibitemOpen
  \bibfield  {author} {\bibinfo {author} {\bibfnamefont {C.}~\bibnamefont
  {Chandrashekar}},\ }\href@noop {} {\bibfield  {journal} {\bibinfo  {journal}
  {arXiv preprint arXiv:1212.5984}\ } (\bibinfo {year} {2012})}\BibitemShut
  {NoStop}%
\bibitem [{\citenamefont {Laine}\ \emph {et~al.}(2010)\citenamefont {Laine},
  \citenamefont {Piilo},\ and\ \citenamefont {Breuer}}]{laine2010measure}%
  \BibitemOpen
  \bibfield  {author} {\bibinfo {author} {\bibfnamefont {E.-M.}\ \bibnamefont
  {Laine}}, \bibinfo {author} {\bibfnamefont {J.}~\bibnamefont {Piilo}}, \ and\
  \bibinfo {author} {\bibfnamefont {H.-P.}\ \bibnamefont {Breuer}},\
  }\href@noop {} {\bibfield  {journal} {\bibinfo  {journal} {Phys. Rev. A}\
  }\textbf {\bibinfo {volume} {81}},\ \bibinfo {pages} {062115} (\bibinfo
  {year} {2010})}\BibitemShut {NoStop}%
\bibitem [{\citenamefont {Breuer}\ \emph {et~al.}(2009)\citenamefont {Breuer},
  \citenamefont {Laine},\ and\ \citenamefont {Piilo}}]{breuer2009measure}%
  \BibitemOpen
  \bibfield  {author} {\bibinfo {author} {\bibfnamefont {H.-P.}\ \bibnamefont
  {Breuer}}, \bibinfo {author} {\bibfnamefont {E.-M.}\ \bibnamefont {Laine}}, \
  and\ \bibinfo {author} {\bibfnamefont {J.}~\bibnamefont {Piilo}},\
  }\href@noop {} {\bibfield  {journal} {\bibinfo  {journal} {Phys. Rev. Lett}\
  }\textbf {\bibinfo {volume} {103}},\ \bibinfo {pages} {210401} (\bibinfo
  {year} {2009})}\BibitemShut {NoStop}%
\bibitem [{\citenamefont {Breuer}\ \emph {et~al.}(2016)\citenamefont {Breuer},
  \citenamefont {Laine}, \citenamefont {Piilo},\ and\ \citenamefont
  {Vacchini}}]{breuer2016colloquium}%
  \BibitemOpen
  \bibfield  {author} {\bibinfo {author} {\bibfnamefont {H.-P.}\ \bibnamefont
  {Breuer}}, \bibinfo {author} {\bibfnamefont {E.-M.}\ \bibnamefont {Laine}},
  \bibinfo {author} {\bibfnamefont {J.}~\bibnamefont {Piilo}}, \ and\ \bibinfo
  {author} {\bibfnamefont {B.}~\bibnamefont {Vacchini}},\ }\href@noop {}
  {\bibfield  {journal} {\bibinfo  {journal} {Rev. Mod. Phys}\ }\textbf
  {\bibinfo {volume} {88}},\ \bibinfo {pages} {021002} (\bibinfo {year}
  {2016})}\BibitemShut {NoStop}%
\bibitem [{\citenamefont {Rivas}\ \emph {et~al.}(2010)\citenamefont {Rivas},
  \citenamefont {Huelga},\ and\ \citenamefont {Plenio}}]{RHP10}%
  \BibitemOpen
  \bibfield  {author} {\bibinfo {author} {\bibfnamefont {{\'A}.}~\bibnamefont
  {Rivas}}, \bibinfo {author} {\bibfnamefont {S.~F.}\ \bibnamefont {Huelga}}, \
  and\ \bibinfo {author} {\bibfnamefont {M.~B.}\ \bibnamefont {Plenio}},\
  }\href@noop {} {\bibfield  {journal} {\bibinfo  {journal} {Phys. Rev. Lett}\
  }\textbf {\bibinfo {volume} {105}},\ \bibinfo {pages} {050403} (\bibinfo
  {year} {2010})}\BibitemShut {NoStop}%
\bibitem [{\citenamefont {Luo}\ \emph {et~al.}(2012)\citenamefont {Luo},
  \citenamefont {Fu},\ and\ \citenamefont {Song}}]{luo2012quantifying}%
  \BibitemOpen
  \bibfield  {author} {\bibinfo {author} {\bibfnamefont {S.}~\bibnamefont
  {Luo}}, \bibinfo {author} {\bibfnamefont {S.}~\bibnamefont {Fu}}, \ and\
  \bibinfo {author} {\bibfnamefont {H.}~\bibnamefont {Song}},\ }\href@noop {}
  {\bibfield  {journal} {\bibinfo  {journal} {Phys. Rev. A}\ }\textbf {\bibinfo
  {volume} {86}},\ \bibinfo {pages} {044101} (\bibinfo {year}
  {2012})}\BibitemShut {NoStop}%
\end{thebibliography}%
\end{document}